\documentclass[preprint,12pt]{elsarticle}
\newcommand{\R}{{\mathbb{R}}}
\newcommand{\Z}{{\mathbb{Z}}}
\usepackage{amssymb}

\journal{Physica E: Low-dimensional Systems and Nanostructures}

\begin{document}
\begin{frontmatter}

\title{  
Origin of resonant tunneling through single-point barriers 
 }
\author{A.V. Zolotaryuk}
\address
{Bogolyubov Institute for Theoretical Physics, National Academy of
Sciences of Ukraine, Kyiv 03143, Ukraine}

\date{\today}

\begin{abstract}
The physical interpretation of the appearance of resonant transmission through
single-point barriers is discussed on the basis of a
 double-layer heterostructure  in the squeezing limit as both 
 the thickness of the layers
and the distance between them tend  to zero simultaneously. 
In this limit, the electron transmission through a barrier-well structure is derived
to be non-zero at certain discrete values of the system parameters forming the so-called
 resonance set, while beyond this set, the structure behaves as a perfectly
reflecting wall. The origin of this phenomenon is shown to result from 
the reflection coefficients at the  interfaces in the inter-layer space. 
The transmission amplitude
is computed as a set function defined on the trihedral angle surface in a 
three-dimensional parameter space.
  \end{abstract}

\begin{keyword} \\
One-dimensional quantum systems \\
 Transmission \\
Point interactions \\  
 Resonant tunneling

\end{keyword}

\end{frontmatter}

\section{Introduction}

Since the pioneering 
 studies  \cite{te,ec,cet} of resonant transmission
through quantum multilayer  heterostructures, 
electronic tunnel systems are a source of considerable
interest. These structures  are not only important in micro- and nanodevices,
but their study involves a great deal of basic physics. In
recent years it has been realized that the study of the electron transmission
through heterostructures can be investigated in the zero-thickness limit
approximation when their width shrinks to zero.
Within such an approximation it is possible 
to produce the so-called  {\em point interaction} models
(see books \cite{do,a-h} for details and references) which are  quite useful because
they admit exact closed analytical
solutions  providing  relatively
simple situations, where an appropriate way of squeezing to zero
 can be chosen to be in relevance with a real structure. 

The present paper focuses on the investigation of the physical mechanism of 
resonant tunneling  through the planar heterostructure composed of
 extremely thin layers separated by some small distances in the limit as
both the layer thickness and the distance between the layers  simultaneously
tend to zero. The electron motion in this system is confined in the longitudinal
direction (say, along the $x$-axis), which is perpendicular to the planes, and is
free in the transverse direction.
The three-dimensional Schr\"{o}dinger equation of such a structure can be separated
into longitudinal and transverse parts, writing the total electron energy
as the sum of the longitudinal and transverse energies: 
$E_l+ \hbar^2 {\bf k}_t^2/2m^*$, where
$m^{*}$ is an effective electron mass and ${\bf k}_t$ the transverse wave vector,
 and expressing the wave function by the
product $\psi  =\psi_l \psi_t$. As a result, we arrive at the reduced one-dimensional
Schr\"{o}dinger equation with respect to the longitudinal component of the
wave function $\psi_l(x)$ and the electron energy $E_l$. For brevity of notations,
in the following we omit the subscript $``l''$ at both  $\psi_l(x)$ and $E_l$. Thus,
in the units as $\hbar^2/2m^*=1$, 
 the one-dimensional stationary Schr\"{o}dinger equation reads
\begin{equation}
-\psi''(x) +V_\varepsilon(x)\psi(x)= E\psi(x),
\label{1}
\end{equation}
where the potential $V_\varepsilon(x)$ is given in terms of piecewise constant functions. 
It is supposed to depend on the squeezing parameter $\varepsilon > 0$, so that
in the limit as $\varepsilon \to 0$, the function $V_\varepsilon(x)$ is confined to
one point. 

A whole body of the mathematical and physical literature (see, e.g., 
\cite{s,k,adk,cnp,cnt,cft,tfc,gnn,l1,bn,gggm,l2,gmmn,f-n}, a few to mention)
has been published where a number of interesting features of
 point interactions was discovered for the one-dimensional Schr\"{o}dinger 
equation with singular potentials in the form of distributions.
In  the work \cite{c-g}, it was suggested to regularize 
 the potential $\gamma \delta'(x)$, 
$\delta'(x) := d\delta(x)/dx$, where $\delta(x)$ is Dirac's delta function and 
$\gamma$ the potential strength (intensity), by a barrier-well potential profile
$V_\varepsilon(x)$, and then perform the $\varepsilon \to 0$ limit. As a result,
a non-zero transmission through this singular single-point barrier 
has been shown to occur 
under certain  conditions imposed on the intensity $\gamma$
forming a discrete resonance set.
 Later on, it has been observed \cite{zci}
that the resonance set for this point potential  depends
on the piecewise constant regularization of the distribution $\delta'(x)$.
Next, this family of point interactions has been 
extended using the $\delta'$-like regularizing sequences of a more
general type including those which are beyond piecewise constant functions \cite{tn}.
Finally, it  has rigorously  been proved  \cite{gm,gh,g1,gh1,g2,g3} the existence of the resonance 
set for the potential $\gamma \delta'(x)$ for {\em arbitrary} $\delta'$-like regularizing 
sequence. Moreover, it has been derived  that this set depends  on the shape of the regularizing potential leading to  the conclusion about the existence of a hidden parameter in the
$\delta'$-potential (see also \cite{zz14}). Basically, e.g., in the works 
 \cite{zz14,zpla10,zz11,z17}, the realization of the one-point 
resonant-tunneling interactions has been treated  as a 
 cancellation of divergences in the squeezing limit. Therefore, it would be of interest
to give the physical interpretation  of the origin of this phenomenon
and this is the main purpose of the present paper.

The procedure of looking for the resonance sets for the point interactions,
 which are realized from multilayer structures,
 can  be described briefly as follows. 
Within each layer the potential $V_\varepsilon(x)$ is constant and therefore Eq.\,(\ref{1})
can easily be solved. The solution can be represented via the transmission matrix
connecting the boundary conditions for the wave function $\psi(x)$ at the left and 
right interfaces of the structure. Let the structure be located on the interval 
$(x_1, x_2)$. Then transmission matrix $\Lambda_\varepsilon$ is defined
by the  equations
\begin{eqnarray}
\left( \begin{array}{cc} \psi(x_2)  \\
\psi'(x_2) \end{array} \right)  = {\Lambda}_\varepsilon \left(
\begin{array}{cc} \psi(x_1)   \\
\psi'(x_1)   \end{array} \right), ~~~ {\Lambda}_\varepsilon = 
 \left( \begin{array}{cc} {\lambda}_{11,\varepsilon}~~ {\lambda}_{12,\varepsilon} \\
{\lambda}_{21,\varepsilon} ~~{\lambda}_{22,\varepsilon} \end{array} \right) ,
\label{1a}
\end{eqnarray}
where the matrix elements $\lambda_{ij,\varepsilon}$ satisfy the relation
\begin{equation}
\lambda_{11,\varepsilon}\lambda_{22,\varepsilon}-
\lambda_{12,\varepsilon}\lambda_{21,\varepsilon} =1,
\label{2a}
\end{equation}
being valid for any $\varepsilon >0$. As usual, the squeezing limit is arranged in such
a way that $x_1 \to -0$ and $x_2 \to +0$  as $\varepsilon \to 0$.

In general, for any multilayer structure the limit relations
 $\lim_{\varepsilon \to 0}\lambda_{12,\varepsilon}=0$ and 
$\lim_{\varepsilon \to 0}|\lambda_{21,\varepsilon}|=\infty$ hold true. Under certain
conditions on the system parameters, 
$\lim_{\varepsilon \to 0}\lambda_{11,\varepsilon}$ and
$\lim_{\varepsilon \to 0}\lambda_{22,\varepsilon}$ can be finite and in this case
the two-sided boundary conditions become of the Dirichlet type: 
$\psi(\pm 0)=0$. In physical terms, 
this means that 
the limit point structure acts as a perfectly reflecting wall.
Since the element $\lambda_{21,\varepsilon}$ is the most singular term in the matrix
$\Lambda_\varepsilon$, one can impose the constraint 
$\lim_{\varepsilon \to 0}\lambda_{21,\varepsilon}=0$ and, if 
this condition is satisfied, it can be viewed as an equation 
on the system parameters. At the parameter values  satisfying this condition,
the so-called  {\em resonance set},
the transmission is non-zero (partial or perfect), while beyond this set 
the point structure is completely opaque. 

In all the previous publications  the 
condition $\lim_{\varepsilon \to 0}\lambda_{21,\varepsilon}=0$ was
 treated as a {\em cancellation of divergences}
in the limit as $\varepsilon \to 0$, but nowhere the explanation of the 
origin of this phenomenon has been undertaken. This paper aims  to 
explain how the resonant tunneling of this type happens in the simplest case
of a double-layer structure. For this purpose we use the interference mechanism,
similarly to that used in the works \cite{knt1,knt2}, where
 instead of the description of point interactions in terms of 
the limiting transmission matrix
 $\Lambda := \lim_{\varepsilon \to 0}\Lambda_\varepsilon $,
an alternative way for identifying the whole family of
point interactions has been used. This approach has been suggested 
in the works \cite{cft,tfc}, according to which
the boundary conditions are written via the two-component vectors
\begin{eqnarray}
 \Psi := \left( \begin{array}{cc} \psi(+\,0)   \\
\psi(-\,0)   \end{array} \right), ~~ 
\Psi' := \left( \begin{array}{cc} \psi'(+\,0)   \\
-\, \psi'(-\,0)   \end{array} \right).
\label{3a}
\end{eqnarray}
The matrix equation for $\Psi$ and $\Psi'$ reads 
\begin{equation}
(U-I)\Psi + {\rm i}L_0 (U+I)\Psi' =0,
\label{4a}
\end{equation}
where $U \in \mbox{U}(2)$ is a two-by-two unitary matrix, $I$ the unit matrix, and
$L_0$ an arbitrary non-zero constant of length dimension.
The $U$-matrix can be parametrized in an appropriate way and the relationship between
its elements and the $\Lambda$-matrix elements can be established 
(for more details see \cite{cft}). Using this approach in \cite{knt1,knt2}, the 
 scattering of a quantum particle by two independent point interactions 
 has been investigated  in one dimension. As a result,  the resonance 
conditions for perfect transmission through this two-point system have been found.

The paper is organized as follows. In Section 2, we define the potential profile
for a double-layer structure and derive the formulae for the  
reflection-transmission coefficients. In the next section, the asymptotic
representation of the resonance condition is obtained in the limit as the structure
shrinks to one point. Based on the power-connecting three-scale parametrization 
of the system parameters, the transmission properties for a whole family of 
single-point interactions are investigated in Section 4.
 Finally, in Section 5, we give the concluding remarks.

\section{Double-layer potential and reflection-transmission coefficients}

We consider the heterostructure composed of two homogeneous layers with 
width $l_1$ and $l_2$ separated at distance $r$. The potential for such a system
can be   expressed as the following
 piecewise constant function:
\begin{equation}
V(x)= \left\{ \begin{array}{ll}
 h_1 &   \mbox{for}~~ x_1 < x < y_1 , \\
 h_2     & \mbox{for}~~ x_2 < x < y_2 ,\\
0 & \mbox{for}~ -\infty < x <x_1 , ~y_1< x< x_2, ~y_2 < x < \infty ,
\end{array} \right. 
\label{2}
\end{equation} 
where $x_1 < y_1 < x_2 < y_2$; $l_1 := y_1 -x_1$, $l_2 := y_2 -x_2$
and $r := x_2 - y_1$.  
The transmission matrix ${\Lambda}_j$ for each layer ($j =1,2$)
 is defined by the relations
\begin{eqnarray}
\left( \begin{array}{cc} \psi(y_j)  \\
\psi'(y_j) \end{array} \right)  = {\Lambda}_j \left(
\begin{array}{cc} \psi(x_j)   \\
\psi'(x_j)   \end{array} \right), ~~~ {\Lambda}_j= 
 \left( \begin{array}{cc} {\lambda}_{j,11}~~ {\lambda}_{j,12} \\
{\lambda}_{j,21} ~~{\lambda}_{j,22} \end{array} \right) .
\label{3}
\end{eqnarray}
Each of these matrices connects the boundary conditions of the wave function
$\psi(x)$ and its derivative $\psi'(x)$ at $x=x_j $ and $x=y_j = x_j + l_j $. 
Explicitly, 
\begin{equation}
  \Lambda_j  =  \left( \begin{array}{lr} ~\cos(k_jl_j)~~ k_j^{-1}\sin(k_jl_j) \\
-\, k_j \sin(k_jl_j)  ~~ \cos(k_jl_j) \end{array} \right) , ~~~k_j := \sqrt{E -h_j}\, ,
~~~j=1,2.
\label{4}
\end{equation}

The scattering coefficients  for each layer 
can directly be expressed through the
elements of the  ${\Lambda}$-matrices (\ref{4}). The reflection and transmission coefficients 
for a quantum particle, incident from the left- and right-hand side and 
 scattered by two layers (denoted by 1 and 2) can be defined by the following relations:
\begin{equation}
\psi(x) = \left\{ \begin{array}{ll}
 {\rm e}^{{\rm i} kx} + R_1^l\, {\rm e}^{-{\rm i} kx} &  \mbox{for}~
-\infty <  x < x_1,  \\
 T_1^l\, {\rm e}^{{\rm i} kx} & \mbox{for}~~~ y_1  < x <  x_2 , 
\end{array} \right. 
\label{5}
\end{equation} 
for the  plane wave $\exp({\rm i}kx)$, incident  from the left-hand side  and 
scattered by layer 1,
\begin{equation}
\psi(x) = \left\{ \begin{array}{ll}
{\rm e}^{{\rm i} kx} + R_2^l\, {\rm e}^{-{\rm i} kx} &  \mbox{for}~~~
y_1 <  x < x_2,  \\
 T_2^l\, {\rm e}^{{\rm i} kx} & \mbox{for}~~~ y_2  < x <  \infty ,
\end{array} \right. 
\label{5a}
\end{equation} 
for the plane wave $\exp({\rm i}kx)$, propagating in the inter-layer space and 
scattered by layer 2, and 
\begin{equation}
\psi(x) = \left\{ \begin{array}{ll}
 {\rm e}^{-{\rm i} kx} + R_2^r\, {\rm e}^{{\rm i} kx} &  \mbox{for}~~~
y_2 <  x < \infty,  \\
T_2^r\, {\rm e}^{-{\rm i} kx} & \mbox{for}~~~ y_1  < x <  x_2 , 
\end{array} \right. 
\label{6}
\end{equation} 
for the  plane wave $\exp(-{\rm i}kx)$, incident  from the right-hand side  and
scattered by layer 2,
\begin{equation}
\psi(x) = \left\{ \begin{array}{ll}
{\rm e}^{-{\rm i} kx} + R_1^r\, {\rm e}^{{\rm i} kx} &  \mbox{for}~~~
y_1 <  x < x_2,  \\ 
 T_1^r\, {\rm e}^{-{\rm i} kx} & \mbox{for}~ - \infty   < x < x_1 ,
\end{array} \right. 
\label{6a}
\end{equation} 
for the  plane wave $\exp(-{\rm i}kx)$, propagating in the inter-layer space and 
scattered by layer 1. 
Inserting the boundary conditions derived from the definition (\ref{5})-(\ref{6a}) 
into the matrix equations (\ref{3}) with Eqs.\,(\ref{4}), 
we find the following representation of the scattering coefficients in terms
of the $\Lambda_j$-matrices:
\begin{eqnarray}
R^l_j &=& -(u_j +{\rm i} v_j) D_j^{-1} {\rm e}^{2{\rm i}k x_j}, ~~ 
R^r_j = (u_j - {\rm i} v_j) D_j^{-1} {\rm e}^{- 2{\rm i}k y_j}, \nonumber \\
T^l_j &= & T_j^r = 2D_j^{-1} {\rm e}^{ {\rm i}k (x_j -y_j)} , 
\label{7}
\end{eqnarray}
where 
\begin{eqnarray}
u_j &= &\lambda_{j,11}- \lambda_{j,22} = 0, ~~~
v_j =  k \lambda_{j,12}+ k^{-1}\lambda_{j,21} =
 \left( {k \over k_j}- {k_j \over k} \right)\sin(k_jl_j), 
\nonumber \\
D_j & = &  \lambda_{j,11}+ \lambda_{j,22} +{\rm i }( k^{-1}\lambda_{j,21} -
k \lambda_{j,12})\nonumber \\
&& ~~~~~~~~~~~~~~~~~=2\cos(k_jl_j)- {\rm i} \left( {k \over k_j } + { k_j \over k}\right)
\sin(k_jl_j) .
\label{8}
\end{eqnarray}

Consider now the  plane wave
$\exp({\rm i}kx)$, incident upon the whole (double-layer) structure from the left 
and thus the interference effect is present. Similarly, we define the scattering coefficients 
for the whole  system located on the interval $(x_1,y_2)$:
\begin{equation}
\psi(x) = \left\{ \begin{array}{ll}
 {\rm e}^{{\rm i} kx} + R^l\, {\rm e}^{-{\rm i} kx} &  \mbox{for}~
-\infty <  x < x_1,  \\
 T^l\, {\rm e}^{{\rm i} kx} & \mbox{for}~~~ y_2  < x <  \infty . \end{array} \right. 
\label{9}
\end{equation} 
Then summing up all the trajectories according to Fig.\,\ref{fig1}, we derive 
 the following relations for the total reflection and transmission coefficients
defined through Eq.\,(\ref{9}):
\begin{figure}
\centerline{\includegraphics[width=1.0\textwidth]{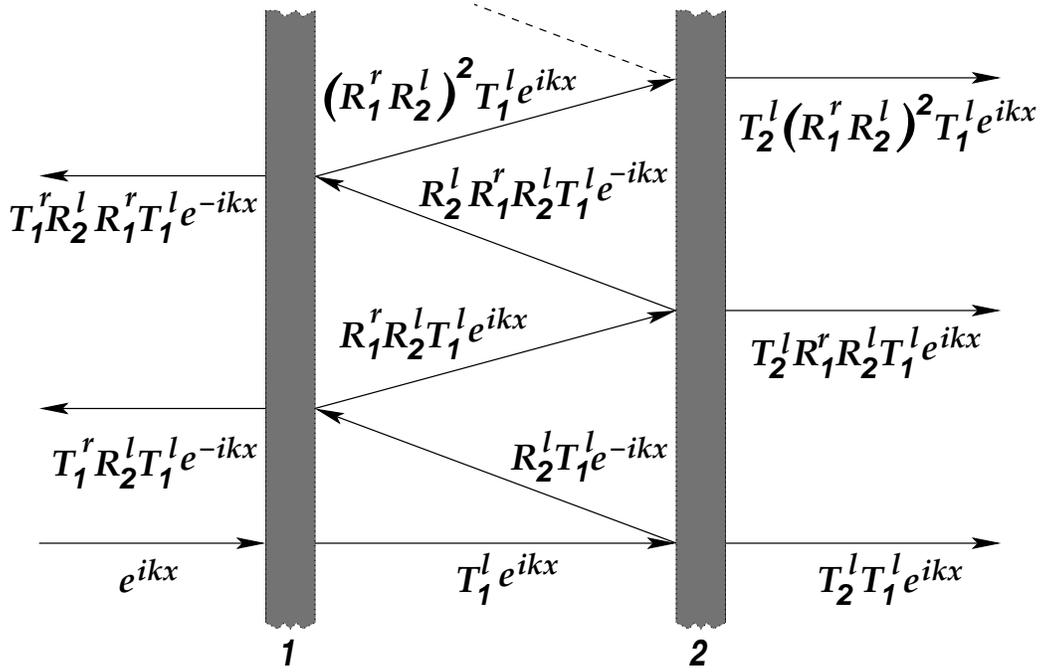}}
\caption{Schematic representation of the first several  reflected and transmitted  
trajectories for the incident plane wave $\exp({\rm i}kx)$
 from the left-hand side, which are 
 scattered by both layers 1 and 2. These trajectories correspond to the 
definition (\ref{5})-(\ref{6a}) and their summing up leads to the formulae (\ref{10}).
}
\label{fig1}
\end{figure}
\begin{equation}
R^l = R_1^l +{ T_1^l T_1^r R_2^l \over 1 - R_1^r R_2^l }~~~\mbox{and}~~~
T^l = { T_1^l T_2^l \over 1 - R_1^r R_2^l }\,.
\label{10}
\end{equation}

\section{Squeezing limit}

Using the explicit representation for the scattering coefficients, which follow from
Eqs.\,(\ref{7}) and (\ref{8}), i.e.,
\begin{eqnarray}
R_j^l &=& {(k/k_j - k_j /k) \exp(2{\rm i}kx_j) \over k/k_j + k_j /k + 2{\rm i}\cot(k_jl_j)}\, , ~~~
R_j^r = {(k/k_j - k_j /k) \exp(-2{\rm i}ky_j) \over k/k_j + k_j /k + 2{\rm i}\cot(k_jl_j)}\, ,
 \nonumber \\
T_j^l & =& T_j^r = {\exp(-{\rm i}kl_j ) \over \cos(k_jl_j) - ({\rm i}/2)
(k/k_j +k_j/k)\sin(k_jl_j)},
\label{11}
\end{eqnarray}
one can estimate separately in the squeezing limit the numerator $T_1^l T_2^l$ and 
the denominator $1- R_1^r R_2^l$ in the second formula (\ref{10}) 
for the total transmission $T^l$. In this limit, we have
 $l_j \to 0$ and $|h_j| \to \infty$, but the products $k_j l_j $, $j=1,2$, as 
the arguments of the trigonometric functions, must be  finite (also including the zero
limit). Hence,
$k_jl_j \to \sqrt{-h_j}\, l_j =: A_j$; $A_j$'s are either real or imaginary, finite or zero.
Therefore $k_j \to A_j/l_j$ and thus $ k/k_j \pm k_j / k \to  \pm  A_j/ kl_j$.
As a result, we obtain the following asymptotics:
\begin{eqnarray}
T_1^lT_2^l & \to & {\exp{[-{\rm i}k(l_1 +l_2)]} \over \cos\!A_1\cos\!A_2\, 
(1 -{\rm i}/2k\alpha_1)( 1- {\rm i}/2k\alpha_2)  }
 = {\cal O}(l_1l_2)\,, ~~~~\label{12a} \\
R_1^r R_2^l & \to & {\exp(2{\rm i}kr)  \over ( 1 +2{\rm i} k  \alpha_1)(
1+  2{\rm i}k  \alpha_2)} \nonumber \\
 &=& \left\{ (1 -4k^2 \alpha_1 \alpha_2)\cos(2kr)
 + 2k (\alpha_1 +\alpha_2 +2{\rm i}k\alpha_1 \alpha_2 )\sin(2kr) \right. \nonumber \\
&& ~~~~~~~ \left. +{\rm i} [2k(\alpha_1 +\alpha_2)\cos(2kr) -\sin(2kr) ] \right\}^{\!-1}, 
\label{12}
\end{eqnarray}
where $\alpha_j := (l_j/A_j)\cot\!A_j \to 0$ as $l_j \to 0$. 
As can be seen from the expression 
(\ref{12a}), the successive transmission through the layers does not depend on
the inter-layer distance $r$ and it completely vanishes in the limit
as  $l_1, l_2 \to 0$. 
Therefore, in general, the whole structure behaves as a fully 
reflecting wall. This is because of 
 the bigger singularity than that of the typical double-delta potential \cite{lc}. 
Such a singularity occurs due to the non-zero finiteness of the arguments $A_j$.

The only possibility for the total transmission $T^l$ to be non-zero
can happen if the denominator $1 - R_1^r R_2^l$ will be of the same
order ${\cal O}(l_1l_2)$ as the
numerator (even in the particular case $r=0$). In general, this is impossible 
because of the  presence of the sum $\alpha_1 + \alpha_2$ in (\ref{12}), so that
the denominator appears to be of the order ${\cal O}(l_1,l_2)$.
However, here one can impose the condition 
\begin{equation}
\sin(2kr) = 2k(\alpha_1 + \alpha_2)\cos(2kr)
\label{12c}
\end{equation}
and then  the expression in the square brackets of (\ref{12}) vanishes. Using next this
equation  in (\ref{12}) once more, we get
\begin{equation}
R_1^r R_2^l \to [ 1 +4k^2 (\alpha_1^2 +\alpha_1 \alpha_2 + \alpha_2^2)
+8{\rm i}k^3 (\alpha_1 + \alpha_2)\alpha_1 \alpha_2 ]\cos(kr).
\label{12b}
\end{equation}  
Thus, if we additionally assume here that $\cos(2kr) \to 1$ or $\sin(kr) \to 0$,
we obtain that $R_1^r R_2^l = 1 + {\cal O}(l_1^2, l_2^2, l_1l_2 )$. Then 
Eq.\,(\ref{12c}) can be rewritten in the simple form:
\begin{equation}
\tan(kr) = k(\alpha_1 + \alpha_2),
\label{12d}
\end{equation}
which can be viewed as the resonance condition on the system parameters
$h_j, l_j, r;\, j=1,2$. The particular case $r=0$ in this equation is also 
appropriate. Therefore the family of all solutions to Eq.\,(\ref{12d}) generates a  discrete
resonance set, at which the transmission amplitude occurs in the form of extremely 
sharp peaks like those shown, e.g., in Fig.\,1 of the work \cite{zz15} as a result of 
 the (non-uniform) pointwise convergence as  $l_1, l_2, \tan(kr) \to 0$. 
In other words, under the resonance condition (\ref{12d}), due to the infinite 
summing of the inter-layer back-forth reflection steps
  $R_1^r R_2^l$, we arrive at the uncertain ratio $0/0$ in the second 
formula (\ref{12}).  This uncertainty should be treated carefully  by computing the full 
expressions for  $R^l$ and $T^l$ and in this way one  can confirm the validity
of the resonance equation (\ref{12d}). 

Thus, inserting Eqs.\,(\ref{11}) into (\ref{10}), we get
\begin{equation}
R^l = - (u +{\rm i}v )D^{-1}\exp{(2{\rm i}kx_1)}, ~~~T^l = 2 D^{-1}\exp{[-{\rm i}(l_1 +l_2 +r)]},
\label{13}
\end{equation}
where
\begin{eqnarray}
 u\!\!\! &= \!\!\! & \left( {k_2 \over k_1} -{k_1 \over k_2}\right)\sin(k_1l_1)\sin(k_2l_2)\cos(kr)
+ \left[ \left( {k \over k_1}-{k_1 \over k}\right)\sin(k_1l_1)\cos(k_2l_2)
\right. \nonumber \\
&& ~~~~~~~~ ~~~~~~~~  ~~~
- \left. \left( {k \over k_2}-{k_2 \over k}\right)\cos(k_1l_1)\sin(k_2l_2)\right]\!\sin(kr),
\label{14} \\
v \!\!\! &= \!\!\! & \left[ \left( {k \over k_1} -{k_1 \over k}\right)\sin(k_1l_1)\cos(k_2l_2)
+  \left( {k \over k_2}-{k_2 \over k}\right)\cos(k_1l_1)\sin(k_2l_2)
\right]\! \cos(kr)  \nonumber \\
&& ~~~~~~~~ ~~~~~~~~  ~~~
+  \left( {k_1k_2 \over k^2}-{k^2 \over k_1 k_2}\right)\sin(k_1l_1)\sin(k_2l_2)\sin(kr),
\label{15} \\
D \!\!\! &= \!\!\! & \left[ 2\cos(k_1l_1)\cos(k_2l_2) -\left( {k_1 \over k_2} +
{k_2 \over k_1}\right) \sin(k_1l_1)\sin(k_2l_2) \right]\! \cos(kr) \nonumber \\
\!\!\! &- \!\!\! &  \left[ \left( {k \over k_1} +{k_1 \over k}\right)
 \sin(k_1l_1)\cos(k_2l_2) + \left( {k \over k_2} +{k_2 \over k}\right)
 \cos(k_1l_1)\sin(k_2l_2)\right]\! \sin(kr) \nonumber \\
\!\!\! & - \!\!\! & {\rm i}\left\{ \left[  \left( {k \over k_1} +{k_1 \over k}\right)
 \sin(k_1l_1)\cos(k_2l_2) + \left( {k \over k_2} +{k_2 \over k}\right)
 \cos(k_1l_1)\sin(k_2l_2)\right]\! \cos(kr)\right. \nonumber \\
\!\!\! & + \!\!\! & \left. \left[ 2 \cos(k_1l_1)\cos(k_2l_2) -
 \left( {k^2 \over k_1 k_2} +{k_1 k_2 \over k^2 } \right)
 \sin(k_1l_1)\sin(k_2l_2)\right]\! \sin(kr)  \right\}.
\label{16}
\end{eqnarray}
 By direct calculations one can prove that $|D|^2 = 4 + u^2 +v^2$ and, as a result, 
the reflection-transmission amplitudes become
\begin{equation}
{\cal R}^l := |R^l|^2 = {u^2 +v^2 \over 4 + u^2 +v^2}~~~\mbox{and}~~~
{\cal T}^l := |T^l|^2 = {4 \over 4 +u^2 +v^2}\,
\label{17}
\end{equation}
fulfilling the conservation law ${\cal R}^l +{\cal T}^l =1$ for the electron flow.
Next, in the limit as $l_1,l_2 \to 0$, we get the following asymptotic representation:
\begin{eqnarray}
u & \to  &
\left[\left( {l_1^2 \over A_1^2  }-{ l_2^2 \over A_2^2}\right)\! \cos(kr)
 +(\alpha_1 - \alpha_2){\sin(kr) \over k}  \right]
{ \cos\!A_1 \cos\!A_2 \over \alpha_1 \alpha_2}\,,\label{18a} \\
v& \to & - \left[ ( \alpha_1 + \alpha_2 )\cos(kr)- {\sin(kr) \over k }\right]
 { \cos\!A_1 \cos\!A_2 \over  k\alpha_1 \alpha_2}\,, 
\label{18b} \\
D  & \to  &
 \left[ \left( 2\alpha_1 \alpha_2 - {l_1^2 \over A_1^2} - {l_2^2 \over A_2^2} \right)\cos(kr)-
(\alpha_1 + \alpha_2) {\sin(kr) \over k} \right. \nonumber \\
&& ~~~~ + \left. {{\rm i} \over k} \left({\sin(kr) \over k }- (\alpha_1 +\alpha_2)
\cos(kr) \right) \right]
{\cos\!A_1 \cos\!A_2 \over \alpha_1 \alpha_2}\,.
\label{18}
\end{eqnarray}
Under the resonance condition (\ref{12d}),
  these asymptotics are finite. If additionally we assume $r \to 0$, 
this condition becomes
\begin{equation}
r = {l_1 \over A_1}\cot\!A_1 + {l_2 \over A_2}\cot\!A_2\, , ~~~~0 \le r < \infty ,
\label{19}
\end{equation}
and, as a result, the
asymptotics (\ref{18a})-(\ref{18})  reduce to the following simple expressions: 
\begin{equation}
u \to  \theta -\theta^{-1} , 
~~~v \to 0, ~~~D \to  \theta + \theta^{-1}, ~~~
\theta : = - \, { A_1 l_2 \sin\!A_1 \over A_2 l_1 \sin\!A_2 }\,.
\label{20}
\end{equation}
The zero inter-layer distance ($r=0$) can be considered as a particular case 
of the resonance condition (\ref{19}).
Clearly, the limiting expressions (\ref{18a}) and (\ref{18b}) 
satisfy Eqs.\,(\ref{17}), retaining 
the conservation law ${\cal R}^l + {\cal T}^l =1$.

One of the important conclusions which immediately follows from the
 resonance equation (\ref{19}) is the impossibility to realize
a single-point interaction from the double-barrier system. Indeed,
for both the barriers we set $A_j ={\rm i}\bar{A}_j$, where $\bar{A}_j 
:= \lim_{h_j \to \infty,\, l_j \to 0}(\sqrt{h_j}\,l_j) >0$, $j=1,2$, so that
Eq.\,(\ref{19}) reduces to 
\begin{equation}
r =-\sum_{j=1,2}(l_j /\bar{A_j})\coth\!A_j .
\label{20a}
\end{equation}
Since $r >0$, whereas the right-hand side of Eq.\,(\ref{20a})  is negative,
 the zero squeezing of the distance between the barriers is forbidden
and this result agrees with the studies in  \cite{z17}. 
 On the other hand, if one of the layers is a well, say the first one,
the resonance equation reads 
\begin{equation}
r = (l_1/A_1)\cot\!A_1 - (l_2/\bar{A}_2)\coth\!\bar{A}_2,
\label{20b}
\end{equation}
where both the terms $A_1$ and $\bar{A}_2$ are positive, so that Eq.\,(\ref{20b})
can be satisfied. Indeed, while varying $A_1$, one can 
examine the existence of a countable number of solutions to this equation. 

One of the effective ways
to analyze a whole family of single-point interactions in the limit as 
the parameters $l_1, l_2$ and $ r $  tend to zero simultaneously is their representation
through a single squeezing parameter $\varepsilon \to 0$ using different 
power-connecting   relations. Below we will investigate the squeezing limit
using a three-scale approach. In this way we generalize the family of point
interactions obtained previously in several papers \cite{c-g,zci,zz14,zpla10,zz11,z17}.

\section{Power-connecting three-scale parametrization}

The resonance condition (\ref{19}) is given in the asymptotic form as 
the layer thickness parameters $l_1, l_2$ and the inter-layer distance $r$
simultaneously shrink to one point. To proceed with the further analysis
of the condition (\ref{19}), one can simplify the one-point limit procedure by
connecting these parameters through a single squeezing parameter $\varepsilon \to 0$.
The natural connection can be done by using different powers of $\varepsilon$. 
Within such an approach, the three-scale parametrization \cite{zz11}
that   connects the layer parameters through the  parameter $\varepsilon > 0$, 
can be used. Thus, we set
\begin{equation}
h_1 =a_1 \varepsilon^{-\mu}, ~~h_2 =a_2 \varepsilon^{-\nu},~~l_1 =\varepsilon,~~
l_2 = \eta \varepsilon^{1-\mu +\nu},~~r =c\varepsilon^\tau ,~~c \ge 0,
\label{21}
\end{equation}
where $a_j \in \R$, $j=1,2$, and $\mu, \nu, \tau, \eta$ are arbitrary positive
parameters. We denote the potential (\ref{2}) parametrized by these relations by
$V_\varepsilon(x)$. Our task is to describe the possible single-point 
interactions, which can be realized from all the limits 
$V_\varepsilon(x) \to \gamma \delta'(x)$ (in the sense of distributions on the 
$C_0^\infty$ test functions). 

The first step is 
 to find the whole set in the $\{ \mu > 0, \, \nu > 0, \, \tau > 0 \}$-octant,  
where the potential $\gamma \delta'(x)$
can be defined in the standard distributional sense. To this end, we need to estimate 
in the  limit as $\varepsilon \to 0$ the integral
\begin{equation}
\langle V_\varepsilon \, | \, \varphi \rangle = \left( a_1 \varepsilon^{-\mu} \int_0^{l_1} 
+ \,\, a_2 \varepsilon^{-\nu} \int_{l_1+r}^{l_1+r+l_2}\, 
\right)\varphi(x)dx
\label{22}
\end{equation}
for any  $\varphi(x) \in C_0^\infty (\R)$.
Using the parametrization (\ref{21}) and expanding  $\varphi(\varepsilon \xi)=
\varphi(0) +\varepsilon \xi \varphi'(0) + (\varepsilon^2 \xi^2 /2)\varphi''(\varsigma \varepsilon
\xi)$, where $\xi = x/\varepsilon$ and  $\varsigma \in (0,1)$ depends for a given
 $\varphi$ on $\varepsilon \xi$, the integral (\ref{22}) can be computed explicitly.
Thus, under the condition $a_1 +\eta a_2 =0 ~( 0< \eta < \infty)$,  we get
\begin{equation}
\langle V_\varepsilon \, | \, \varphi \rangle  = - \, a_1
\left( \frac12 \varepsilon^{2-\mu} + {\eta \over 2}
\varepsilon^{2(1-\mu) +\nu} + c \varepsilon^{1-\mu +\tau}\right)\varphi'(0) + R_2 
\label{23}
\end{equation}
where  the last term can be estimated as follows
\begin{eqnarray}
| R_2| &\le & \max_{\xi \in \R}\varphi''(\xi)|a_1|\left( {1 \over 3}\varepsilon^{3-\mu} + 
{\eta^2 \over 6}\varepsilon^{3(1-\mu) +2\nu} + {\eta \over 2}\varepsilon^{3-2\mu +\nu}\right.
\nonumber \\
&& \left. + {\eta c \over 2}\varepsilon^{2(1-\mu)+\nu +\tau} +c\, \varepsilon^{2-\mu +\tau }
+ {c^2 \over 2}\varepsilon^{1-\mu +2\tau} \right) .
\label{24}
\end{eqnarray}
Under the inequalities 
\begin{equation}
1 < \mu \le 2,~~~2(\mu -1) \le \nu < \infty,~~~\mu -1 \le \tau < \infty, 
\label{25}
\end{equation}
it is easy to  be convinced that all the powers of $\varepsilon$
[for the terms in the brackets of (\ref{24})] are positive. 
Therefore $R_2 \to 0$ as $\varepsilon \to 0$ and the 
limit $V_\varepsilon(x) \to \gamma \delta'(x)$ (in the sense of distributions),
where the constant $\gamma \in \R$ is the intensity of the $\delta'$-potential, 
leads to the relations 
\begin{equation}
a_1 = 2\gamma/\zeta_Q ~~~~ \mbox{and}~~~~ a_2 = - 2\gamma/\eta \zeta_Q
\label{25a}
\end{equation}
 with  the set function 
\begin{equation}
\zeta_Q = \lim_{\varepsilon \to 0}\left(  \varepsilon^{2-\mu} + \eta 
\varepsilon^{2(1-\mu) +\nu} + 2c \varepsilon^{1-\mu +\tau}\right) =
 \left\{ \begin{array}{lllllll}
 1+\eta  + 2c   &   \mbox{at}~P, \\
 \eta  + 2c & \mbox{on}~K, \\
1 +2c & \mbox{on}~L, \\
1 +\eta & \mbox{on}~N, \\
\eta & \mbox{on}~X, \\
1 & \mbox{on}~Y, \\
2c  & \mbox{on}~Z. \\
\end{array} \right. 
\label{26}
\end{equation}
\begin{figure}
\centerline{\includegraphics[width=1.0\textwidth]{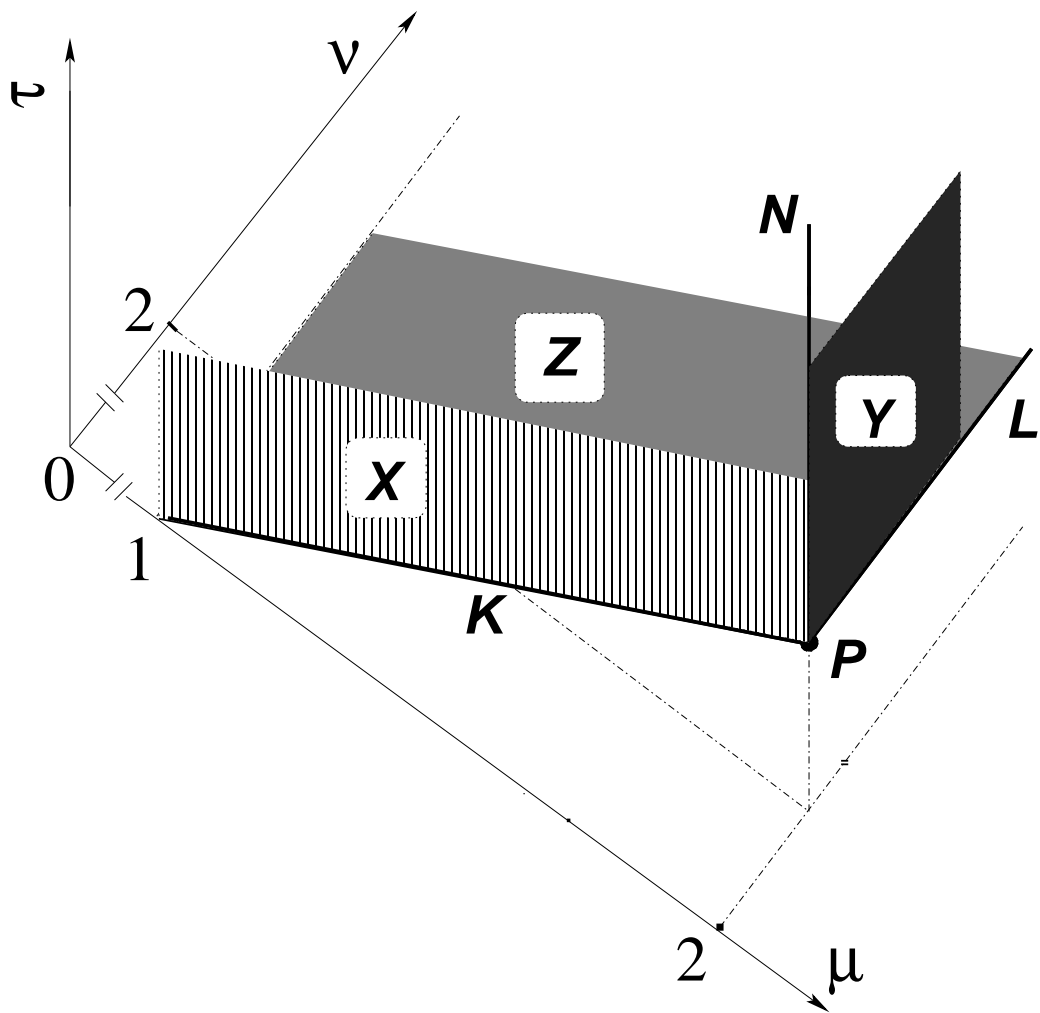}}
\caption{
Trihedral angle surface $S_{\delta'}  = P\cup K \cup L \cup N \cup  X \cup Y \cup Z$
 formed by vertex $P$, three edges $K,L,N$
and three planes $X,Y,Z$, which are defined by Eqs.\,(\ref{27}).
}
\label{fig2}
\end{figure}
Here $Q= P, K,L,N, X,Y,Z$ are the sets in the $(\mu,\nu, \tau)$-octant 
defined by
\begin{eqnarray}
\begin{array}{lllllll}
\mbox{vertex}~P &:=& \{ \mu =\nu=2,\, \tau =1 \},  \\
\mbox{edge}~K & := & \{ 1< \mu < 2, \, \nu= 2(\mu -1), \, \tau= \mu -1 \},  \\
\mbox{edge}~L & :=& \{ \mu =2, \, 2 < \nu < \infty, \, \tau =1 \}, \\
\mbox{edge}~N &:=& \{ \mu =\nu =2, \, 1 < \tau < \infty \},  \\
\mbox{plane}~X &:=& \{ 1 < \mu < 2, \, \nu=2(\mu -1), \, \mu -1 < \tau < \infty \} , \\
 \mbox{plane}~Y &:=& \{  \mu = 2,\,  2 < \nu < \infty, \, 1 < \tau < \infty \} , \\
 \mbox{plane}~Z & : =& \{ 1 < \mu < 2,\, 2(\mu -1) < \nu < \infty, \, \tau =\mu-1 \},
\label{27}
\end{array}
\end{eqnarray}
and  forming the trihedral angle surface $S_{\delta'}$ as shown in Fig.\,\ref{fig2}. 
Note that the inequalities (\ref{25}) are fulfilled on this surface. According to 
Eqs.\,(\ref{25a}),  
on the plane $Z$, the distribution $\gamma \delta'(x)$ makes sense only if $c >0$.

In general, on the whole $S_{\delta'}$-surface, the resonance condition (\ref{19})
parametrized by Eqs.\,(\ref{21}) with the amplitudes (\ref{25a}) becomes an equation
 with respect to the intensity $\gamma$.  
Asymptotically, in the limit as $\varepsilon \to 0$, it takes the form
\begin{eqnarray}
&&  \sqrt{\eta}\, \varepsilon^{\nu/2}
\cot\!\left(\sqrt{2\gamma \eta/\zeta_Q}\,\, \varepsilon^{1-\mu +\nu/2} \right)-
\varepsilon^{\mu/2}\coth\!\left(\sqrt{2\gamma/ \zeta_Q}\, \varepsilon^{1-\mu/2}\right)
 \nonumber \\
&& ~~~~~~~~~~~~~~~~~~~~~~~~~~~
 = c\, \varepsilon^\tau \sqrt{2\gamma/ \zeta_Q}\, .
\label{28}
\end{eqnarray}
Particularly, at the vertex $P$, this equation reduces to 
\begin{equation}
\sqrt{\eta}\, \cot\!\sqrt{2\eta \gamma  \over 1+\eta +2c}=
\coth\!\sqrt{2\gamma \over 1+\eta +2c}+c\,\sqrt{2\gamma \over 1+\eta +2c}
\label{29} 
\end{equation}
and for the case $c=0$ we have the equation, which was derived in \cite{zci}, i.e.,
\begin{equation}
\tan\sqrt{{2\gamma \eta \over 1 +\eta} }=\sqrt{\gamma}\tanh\sqrt{{2\gamma \over 1+\eta }}\, .
\label{30}
\end{equation}
 If, additionally,  $\eta =1$, we arrive at 
the most simple equation $\tan\!\sqrt\gamma = \tanh\!\sqrt\gamma$,
 which was originally obtained in \cite{c-g}.  Clearly, 
 these versions admit a countable 
number of roots $ \gamma_{P,n},~n \in \Z $,  forming the resonance set 
$\Sigma_{P}(\eta,c) := \cup_{n =-\infty}^\infty \gamma_{P,n}$.

For the edges $K$ and $L$, in the limit as $\varepsilon \to 0$ Eq.\,(\ref{28}) 
reduces to the following two equations:
\begin{equation}
 \sqrt{\eta}\, \cot\!\sqrt{2\gamma \eta \over \eta +2c}=
\sqrt{ \eta +2c \over 2\gamma }+c\, \sqrt{2\gamma \over \eta +2c}\,, 
\label{31}
\end{equation}
\begin{equation}
  \coth\!\sqrt{ 2\gamma \over 1 +2c}=
\sqrt{ 1 +2c \over  2\gamma } -  c\, \sqrt{ 2\gamma \over 1 +2c}\,,
 \label{32} 
\end{equation}
respectively. Equation (\ref{31}) admits a countable number of solutions 
on the positive half-axis: $0 \le \gamma_{K,n} < \infty$, while Eq.\,(\ref{32}) on the negative
half-axis $- \infty < \gamma_{L,n} \le 0$. The resonance sets are 
$\Sigma_K(\eta,c) := \cup_{n=0}^\infty \gamma_{K,n}$ and 
$\Sigma_L (c):= \cup_{n=0}^\infty \gamma_{L,n}$,
respectively. Particularly, the resonance sets $\Sigma_K(\eta=1, c=0)$ and 
$\Sigma_L(c=0)$ are given by the roots of the simple equations 
\begin{equation}
\tan\!\sqrt{2\gamma} = \sqrt{2\gamma}~~~\mbox{and}~~~
\tanh\!\sqrt{2\gamma} = \sqrt{2\gamma},
\label{33}
\end{equation}
respectively, found previously in \cite{zz14}. On the $Z$-plane, Eq.\,(\ref{28}) is
fulfilled only for $c=0$, but in this case Eqs.\,(\ref{25a}) do not provide the
$V_\varepsilon(x) \to \gamma \delta'(x)$ limit.

On the edge $N$, Eq.\,(\ref{28}) reduces to (\ref{30}), so that
$\Sigma_P (\eta, c=0) = \Sigma_N (\eta) := \cup_{n = -\infty}^\infty \gamma_{N,n}$.
Finally, on the planes $X$ and $Y$, the resonance sets $\Sigma_X$ and $\Sigma_Y$
are given by the roots of Eqs.\,(\ref{33}), respectively, so that 
$\gamma_{X,n} = \gamma_{K,n}(\eta=1, c=0)$ and $\gamma_{Y,n} = \gamma_{L,n}(c=0)$. 
Concerning the plane $Z$, where the distribution $\delta'(x)$ is well defined,
Eq.\,(\ref{28}) for $Q=Z$ does not allow solutions except for $c=0$, but for this case
the limit $V(x) \to \gamma \delta'(x)$ cannot be defined.

Now the resonance condition (\ref{28}) and its explicit representation given by 
Eqs.\,(\ref{29})-(\ref{33}) can be used to compute the transmission amplitude
${\cal T}^l$ using Eqs.\,(\ref{17}) and (\ref{20}). Thus, the transmission amplitude as a set
function of $Q$ and the $n$th resonance level can be rewritten in the form
\begin{equation}
{\cal T}^l_{Q,n} = { 4 \theta_{Q,n}^2 \over  \left( 1 + \theta_{Q,n}^2  \right)^{2}}
\label{34}
\end{equation}
with the asymptotics
\begin{eqnarray}
\theta^2_{Q,n} &\to & \eta \, \varepsilon^{\nu -\mu} 
\sinh^2\!\left(\! \sqrt{2\gamma_{Q,n}/ \zeta_Q}\, \varepsilon^{1-\mu/2}\right) /
\sin^2\!\left(\! \sqrt{2\gamma_{Q,n}\eta / \zeta_Q}\, \varepsilon^{1-\mu +\nu/2}\right) 
\nonumber \\
& =&  \left(\cosh\!B_{Q,n}
+ c \, \varepsilon^{\tau- 1} B_{Q,n} \sinh\!B_{Q,n}\right)^2
+ \eta \varepsilon^{\nu -\mu}\sinh^2\!\!B_{Q,n}, \nonumber \\
&& ~~~~~~~B_{Q,n}:= \sqrt{2\gamma_{Q,n} / \zeta_Q}\, \varepsilon^{1-\mu/2},
~~~~\mbox{as}~\varepsilon \to 0,
 \label{35} 
\end{eqnarray}
where Eq.\,(\ref{28}) has been used.
Explicitly, for each set $Q=P,K,L,N,X,Y$ and the $n$th resonance, we have
\begin{eqnarray}
 \begin{array}{llllll}
\theta_{P,n}^2 &=& \left( \cosh\!\sqrt{2\gamma_{P,n} /\zeta_P}
 + c\sqrt{2\gamma_{P,n}/\zeta_P}\sinh\!\sqrt{2\gamma_{P,n}/\zeta_P}\right)^2\! \\
&& ~~~~~~~~~~~~~ ~~~~~~~+\eta \sinh^2\!\sqrt{2\gamma_{P,n}/\zeta_P}\,,  \\
 \theta_{K,n}^2 &=&  \left( 1 + 2c\gamma_{K,n}/\zeta_K\right)^2\!
 +2\eta \gamma_{K,n} /\zeta_K , \\
\theta_{L,n}^2 &= & \left( \cosh\!\sqrt{2\gamma_{L,n} /\zeta_L } + 
c \sqrt{2\gamma_{L,n} /\zeta_L } \sinh\!\sqrt{2\gamma_{L,n} /\zeta_L }\right)^2\!, \\
\theta_{N,n}^2 &=&  \cosh^2\!\!\sqrt{2\gamma_{N,n} / \zeta_N } 
+ \eta \sinh^2\!\!\sqrt{2\gamma_{N,n} / \zeta_N }\,, \\
 \theta_{X,n}^2 &=& 1 +2\gamma_{X,n}\,, ~~~~~~~~
\theta_{Y,n}^2 = \left( 1 - 2\gamma_{Y,n} \right)^{-1}.  
\label{36}
\end{array} 
\end{eqnarray}

Particularly, using Eq.\,(\ref{34}) and the first formula in (\ref{36}), for $Q=P$  we obtain
the following expression for the transmission amplitude at $c=0$:
\begin{equation}
{\cal T}^l_{P,n}(c=0)= { (1 - \tanh^2\!\chi_n)(1 +\eta \tanh^2\!\chi_n) \over 
\left( 1 + {\eta -1 \over 2}\tanh^2\!\chi_n \right)^2  }\,,
~~\chi_n := \sqrt{2\gamma_{P,n}(c=0) \over 1+\eta}\,.
\label{37}
\end{equation}  
If additionally we set $\eta =1$, this equation reduces to the formula obtained in 
\cite{c-g} with the resonance set $\{ \gamma_{P,n}(\eta=1,c=0)\}_{n=-\infty}^\infty$
satisfying the resonance condition $\tan\!\sqrt{\gamma} =\tanh\!\sqrt{\gamma},~
\gamma \in \R$.

\section{Concluding remarks}

Thus, the heterostructure consisting of two planar homogeneous layers has been 
investigated in the limit as their thickness parameters $l_1$ and $l_2$ 
 tend to zero.  As a result of this squeezing procedure, the asymptotic 
 resonance condition (\ref{12d}) has been derived in a quite general form.  
Under this condition, the  transmission through a double-layer structure has been observed
to be non-zero at certain discrete values of the system parameters forming the so-called
 resonance set, while beyond this set, the structure behaves as a perfectly
reflecting wall.
Because of $l_1, l_2 \to 0$, the limit $\tan(kr) \to 0$ must be accomplished 
as well. The particular case $\tan(kr) =0$ is also appropriate to satisfy the
resonance condition. In other words, for the resonant tunneling to occur,
the inter-layer distance must  shrink sufficiently fast compared with
the squeezing of the layer thickness. 

In the case when $kr$ is found in the neighborhood of any point $n\pi$,
$n=1, 2, \ldots$, we deal with a two-point system. Then the condition
(\ref{12d}) can be satisfied even for a typical double-barrier system
if $kr < n\pi$ in this neighborhood. In this paper, we restrict ourselves to the limit case
$r \to 0$, for which the resonance condition is asymptotically given by Eq.\,(\ref{19}).
The origin of the resonant tunneling  in a squeezed double-layer
heterostructure  results from the requirement that each  reflection step $R_1^rR_2^l$
 at the  interfaces in the inter-layer space must be of the order
$1 + {\cal O}(l_1^2, l_2^2, l_1l_2)$. This requirement provides in the squeezing limit the 
resonant-tunneling penetration through   a barrier-well or a double-well system,
but not for a double-barrier one.

Finally, using the three-scale parametrization (\ref{21}), the transmission amplitude
has been calculated as a set function defined on the trihedral angle surface shown in 
Fig.\,\ref{fig2}, where the potential $\gamma \delta'(x)$ is defined in the sense
of distributions. These calculations generalize the results derived in the 
previous publications.

\bigskip 
{\bf  Acknowledgments}
\bigskip

The author acknowledges the partial financial support from 
the National Academy of Sciences of Ukraine (Project No.~0117U000238).
He thanks Yaroslav Zolotaryuk for valuable suggestions and 
 critical reading the manuscript.

\end{document}